# China leads scientific trends; the West launches new ones

## Nations lead science differently, by amplifying trends or defying them, with lasting consequences for technological competition.

**Authors:** Jeffrey W. Lockhart[1,2], Jamshid Sourati[2,3], Feng Shi[2,4], James Evans[4,5,6]*

[1] Department of Sociology, University of California, Berkeley, CA 94720.
[2] Knowledge Lab, University of Chicago, 5735 South Ellis Avenue, Chicago, IL 60637.
[3] School of Computing, DePaul University, Chicago, IL 60614.
[4] Department of Sociology, University of Chicago, 1126 E 59th St, Chicago, IL 60637.
[5] Santa Fe Institute, 1399 Hyde Park Road, Santa Fe, NM, 87501.

How nations shape the scientific frontier matters for technological competition, but standard metrics, including publication counts, citations, and disruption indices, look backward and fail to distinguish between fundamentally different leadership strategies. We develop two forward-looking model-based measures and apply them to tens of millions of articles since 1990. The first embeds research pathways within an evolving hypergraph of concepts and scientists to identify leadership in emerging areas—work that anticipates where the scientific crowd is heading. The second embeds evolving samples of ideas and disciplines drawn upon in past research to identify leadership in surprising new directions as unexpected combinations become routine and science reorganizes around them. China became the global leader in emerging areas roughly a decade ago, well before it led in volume, reflecting a capacity to detect and amplify nascent consensus at scale. The United States and Europe show the opposite profile: declining emergence shares but persistent leadership in prescient work, especially research bridging disciplinary boundaries. These patterns replicate across databases, attribution methods, and strategic domains, including AI, biotechnology, energy, and semiconductors. Nations lead science by reading the landscape or by reshaping it, and the institutional requirements for each strategy lie in tension. The distribution of these strategies promises to shape the global structure of technological leadership for decades.

* Corresponding author. E-mail: jevans@uchicago.edu (J.E.)

Scientific and technological developments unfold as global, collective phenomena. New ideas and tools accumulate and diffuse as public goods (*1–3*), replacing prior approaches in waves of creative destruction (*4–6*). Yet science funding occurs disproportionately within nations ((*7–9*)), justified by contributions to industrial or military advantage over shorter time scales. This tension focuses attention on national leadership in global science as a proxy for economic and military power (*10–12*). But standard metrics, including publication counts, scientist numbers, and citations, reflect leadership in the past, not future.

We develop two complementary measures of leadership that capture how nations shape scientific trajectories (*13*, *14*). The first identifies research that leads collective attention into emerging areas. This is science that anticipates and amplifies the momentum of the global scientific crowd. The second identifies prescient advances that fork from predictable trajectories, making surprising contributions that prefigure later research. This is science that pivots toward unexpected possibilities, with fields following.

Applying these measures to tens of millions of research articles since 1990, we find China emerged as the global leader in emerging scientific areas a decade ago, well before it led in publication volume, revealing a deliberate strategy to amplify nascent consensus. Yet China, Japan, and Korea lag significantly in prescient advances. The United States and Europe show the opposite pattern: declining shares in emergent trajectories, but persistent leadership in unpredictable breakthroughs.

To operationalize these dimensions, we evaluate national contributions across all scientific areas and within critical strategic domains, including artificial intelligence, biotechnology, energy, and semiconductors. Our first approach identifies leadership in emerging areas by embedding research pathways within an evolving hypergraph of ideas and scientists. In prior work, this model enabled accurate predictions of future discoveries and their discoverers across the sciences (*15*). It reveals "tectonic movements" among scientific concepts and predicts the emergence of new areas as component ideas catalyze. We identify emerging areas where research ideas converge, growth in associated work accelerates, and attention has become substantial. Because scientists are naturally situated within the embedding, we can characterize which countries are poised to lead as a function of their representation among the scientists closest to emergent field epicenters. Nations leading in emerging areas follow trends to anticipate the scientific crowd.

Our second approach identifies leadership by avoiding the crowd. It assesses the probability of each concept combination by projecting it into an evolving embedding of scientific work. Prior research shows that surprising papers are more likely to become high-impact (*16*). We extend this to identify which countries produce prescient papers—work that surprises at publication but becomes routine in future research. Where emergence captures leadership in areas predicted to coalesce, prescience captures leadership that redirects science toward unexpected directions (Fig. 1).

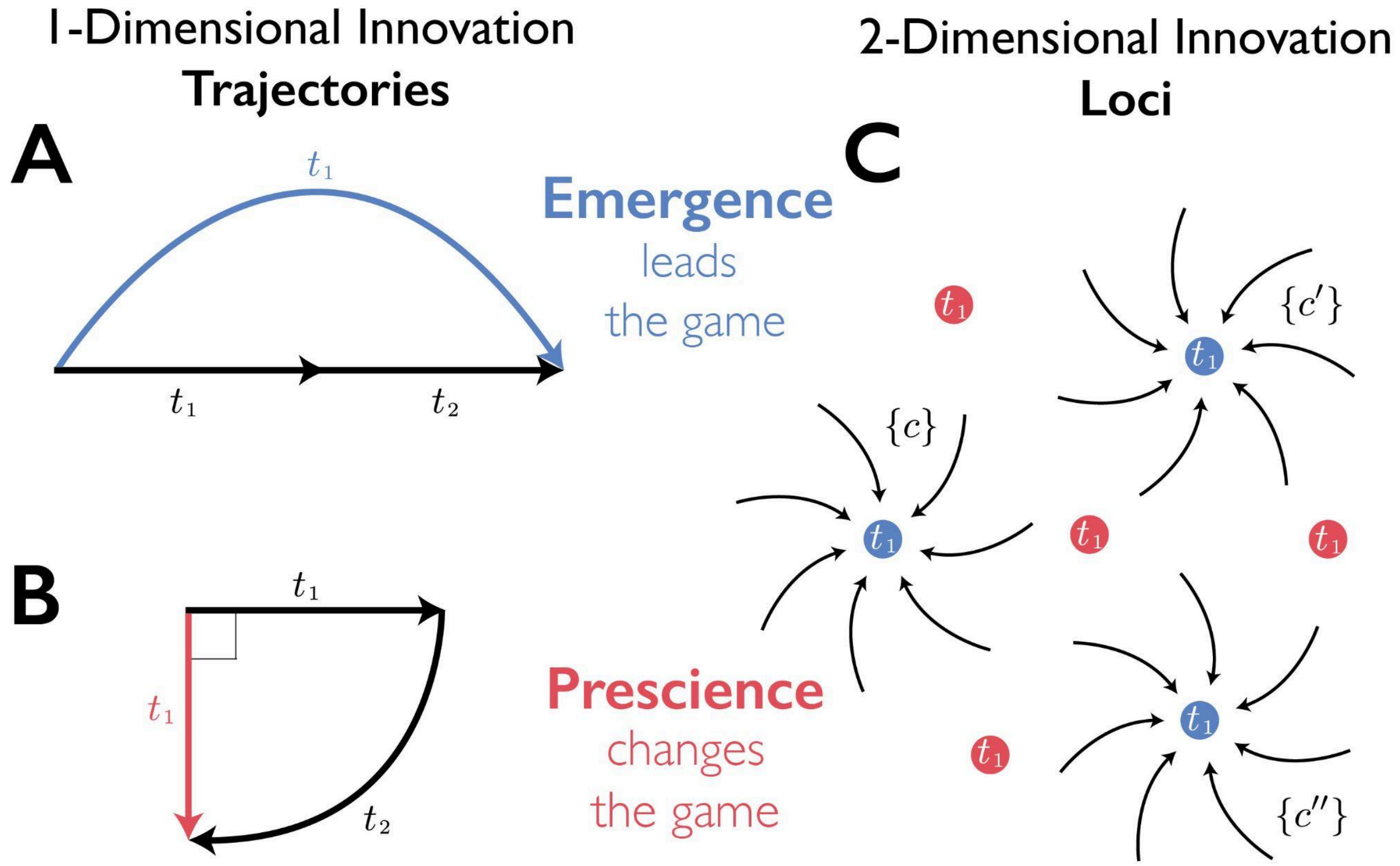

**Figure 1. Conceptual contrast between emergence and prescience approaches to innovation. A-B**, 1-dimensional innovation trajectories where black arrows capture the modal direction of innovation at times 1 and 2. Emergent approaches to innovation at time 1 lead the trajectory of development in time 2 by jumping to the approaching scientific frontier (**A**). Prescient approaches to innovation at time 1 change the trajectory of development by pivoting away from it, and the frontier changes course to meet the prescient innovation (**B**). **C**, 2-dimensional innovation loci at time 1. Sets of concepts, {c}, {c'}, and {c''}, are each converging at time 1. Emergent approaches to innovation anticipate these convergences and prescient approaches pivot away from them. In period 2, concepts across these sets change direction to approach prescient contribution from period 1.

We contrast these forward-looking measures with established metrics: productivity (publication volume), attention (citations) (*17*), and disruption (*18*, *19*). For disruption, we adopt the citation-based disruption index, which quantifies whether a publication eclipses prior work in the citation network (*18*–*21*). It captures whether the community characterizes the focal work as a "brick in the wall" of science by referencing its predecessors or a new foundation by ignoring them. Unlike emergence, which looks prospectively at work that leads to new advances, disruption identifies work credited downstream retrospectively to find pivots in scientific attention.

We demonstrate robustness by replicating analyses across two databases (Web of Science and OpenAlex; Fig. S4), varying paper-to-country attribution methods (Fig. S7), and recomputing models excluding focal countries to confirm global rather than merely local importance (Fig. S5). Innovation areas are globally structured as a shared intellectual community rather than confined to individual nations. While countries invest unequally across areas, no country's advances remain within its borders. Pathbreaking work in one nation spurs global development.

# Results

## Emergence

In our model, when scientific concepts approach within a cosine distance of 0.1, their likelihood of co-occurrence in new research increases by more than 800%. These eruptions of emergent attention define shifts in the landscape of discovery. We capture them by identifying rapidly convergent clusters. These represent expected developments insofar as they are extensions of prior movements along this same landscape (see Methods). Scientists located within the epicenters of these erupting regions are those most consistently and actively engaged in publishing discoveries and inventions there. Scientists located in convergent regions likely to erupt in the near future are poised to lead advances within those areas.

Scientists closest to the epicenter of emerging research areas are disproportionately located at Chinese institutions. Relative to U.S. scientists, Chinese scientists appear in nearly three times as many emergent areas, with four times as many scientists in each. Given the rapid growth of the Chinese academic workforce, it is unsurprising that the share of Chinese authors in emerging areas is also rising. Nevertheless, the timing of the trend is dramatic. While the number of Chinese authors approached those in the U.S. in 2016 and the E.U. in 2019, the representation of Chinese authors among those in emerging areas surpassed those regions 5 years earlier.

**Figure 2. Emergent innovation.** **A**, 2-dimensional UMAP projection of concepts from 100 dimensional embedding of random walks through the hypergraph of papers connected by concepts and authors published in 2020. Embedding these random walks accurately simulates the most likely connection of concepts in future science by existing scientists. Each concept dot is colored by labeled field (e.g., political science), with black bordered dots representing emergent areas, defined according to the criteria pictured in the upper right inset: (1) nearby concept convergence, (2) acceleration of growth in papers with the concept, and (3) a threshold number of papers with the concept. A sample of emergent areas is labeled without parentheses (e.g., *artificial intelligence*, *COVID-19*), and nearby non-emergent areas (e.g., *classification*, *pituitary*) are labeled with parentheses. **B**, Global shares of emergent science from 1990 until 2023 are calculated and credited to the countries with scientists among the 10 closest to the concept area itself and whose work includes that concept. Only the top seven shareholding countries or regions are included in the panel, but all are represented in Figs. S4–S7. The U.S. and E.U. decrease in global share of emergence, while China dramatically increases its share to leadership. The inset contrasts global shares with numbers of emergent papers, which grow for all countries. **C**, Per paper rate of emergence shows this approach has long characterized Japanese and Korean innovation strategies, but China jumps to per-paper emergent leadership over the U.S. in 2016 and over any country or region by 2018 (see Figs. S4–S7 for alternative calculations and robustness checks).

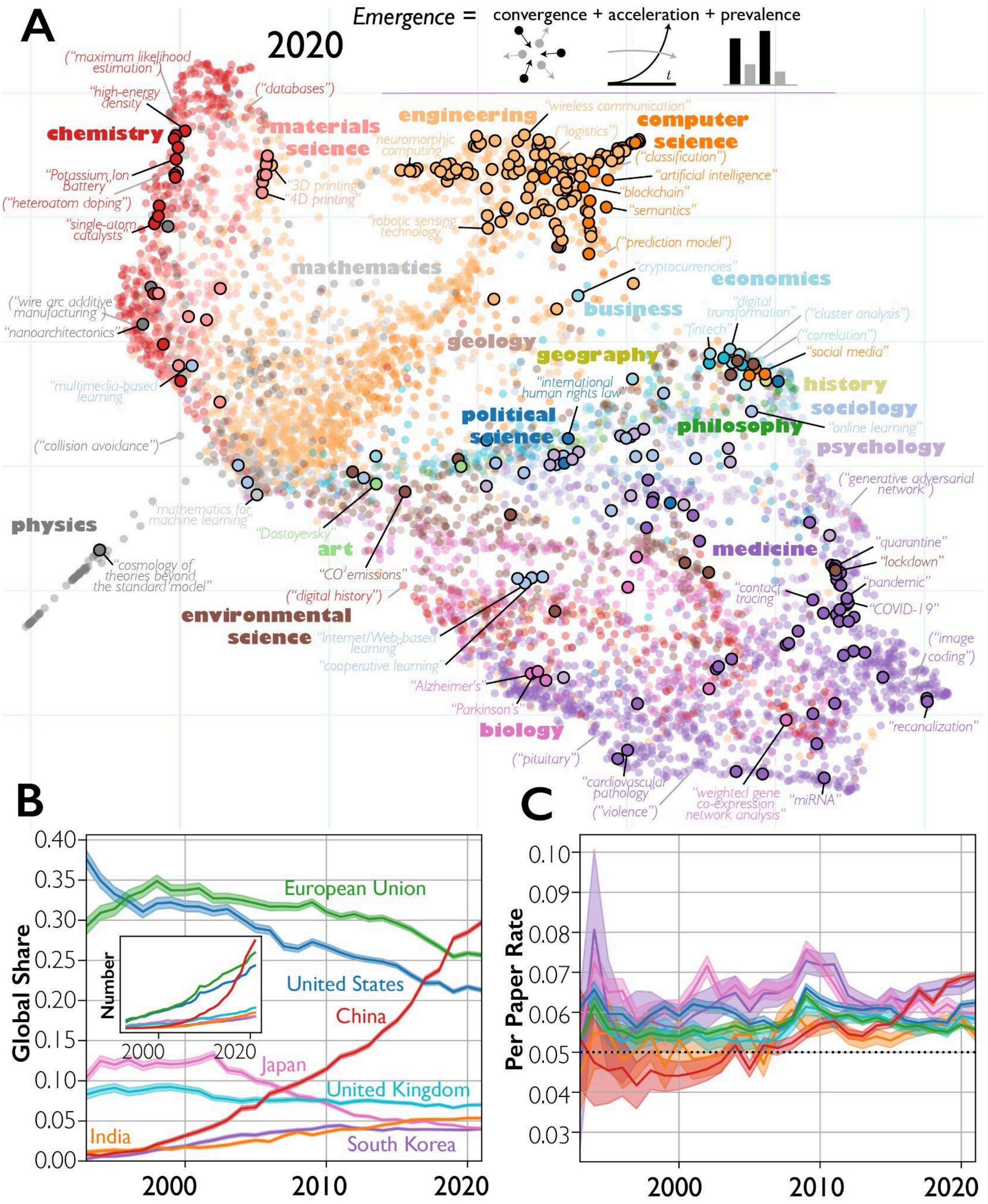

## Prescience

In our dynamic model distinguishing predictable from unpredictable techno-scientific near-term discoveries and inventions, we can identify *prescient* papers, those 'ahead of their time'. These are research products that are highly *un*predictable when published, but later appear highly predictable as the distribution of research pivots and reorganizes around them. In effect, future research follows their lead.

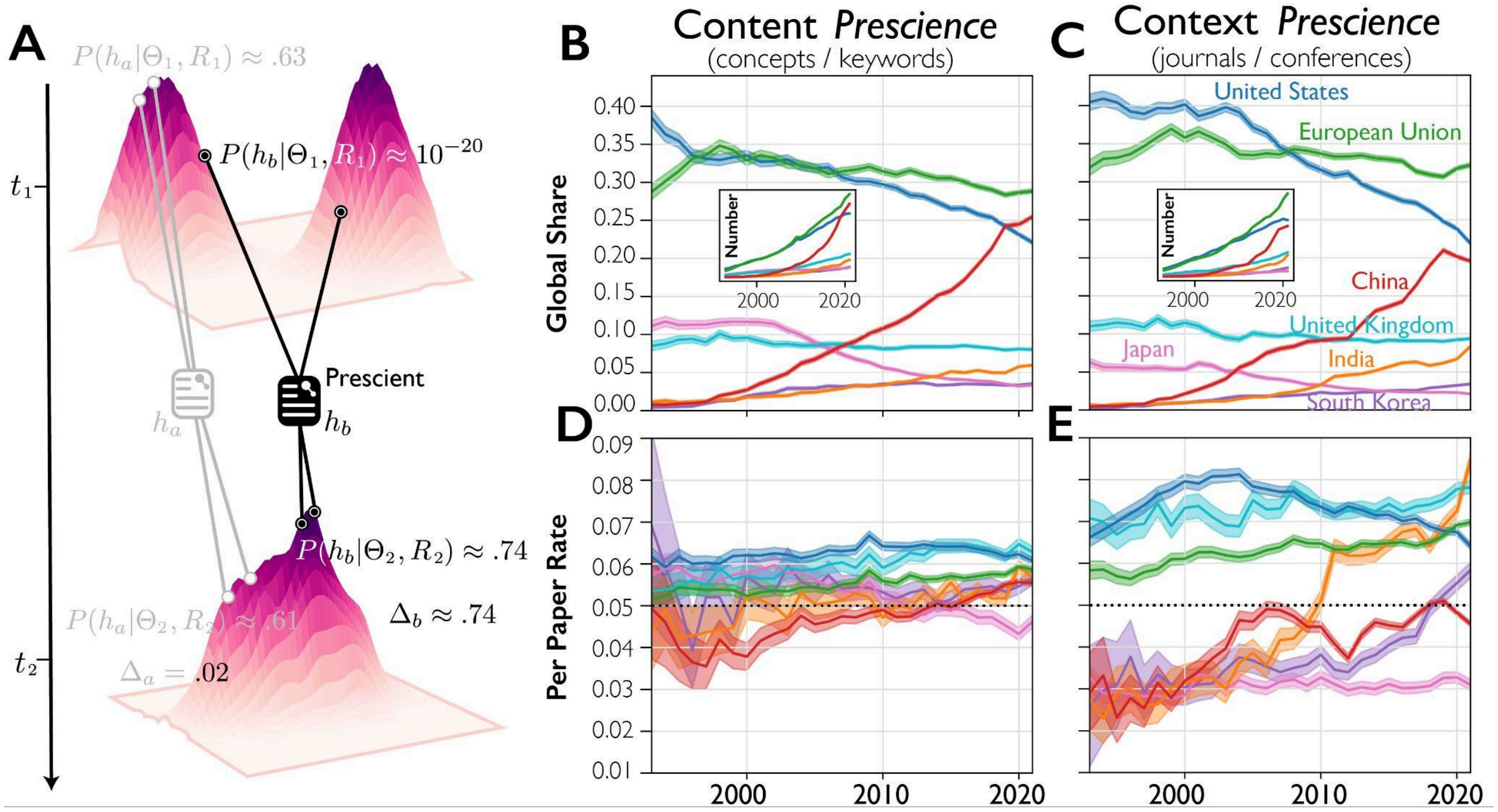

**Figure 3. Prescient innovation. A**, Conceptual illustration of how content and context prescience are calculated and the phenomena they represent. Paper $h_a$ (represented as a hypergraph) combines concepts proximate in the content manifold ($\Theta_1$) as a function of their frequent combination in contemporary papers, and their visibility to contemporary scientists by their frequency in the literature ($R_1$ and proximity to the frequency peak). As such, $h_a$ is unsurprising when published in $t_1$ ($p$ = .63), remaining obvious in $t_2$ ($p$ = .61). By contrast, paper $h_b$ combines less visible ($R_1$) and more distant concepts in ($\Theta_1$), resulting in a very surprising paper ($p = 10^{-20}$). Nevertheless, in period 2, the rest of science has changed, such that the concepts involved in $h_b$ have become unsurprising, now frequent and routinely combined (i.e., the peaks have merged). In terms of its contents, when first published $h_a$ was obvious, but $h_b$ was prescient, anticipating or effecting a change in the game of science. The same figure can be reinterpreted to illustrate context prescience, where $h_a$ combines ideas from obviously related venues, but $h_b$ draws ideas from venues rarely, if ever, used together before. **B-C**, Global shares of prescient science from 1990 until 2023 are calculated and credited to the countries according to their share of top 10% prescient papers, with insets tracing the number of such papers. China's share of prescient papers is growing, leading the world in content prescience but lagging in context prescience. **D-E**, Per paper rates of prescient science reveal a different story. Top Asian countries are somewhat lower than the rest of the world in per-paper content prescience, suggesting a lower likelihood of combining unusual ideas, but China has risen from the bottom to the top of the Asian pack. The same countries are much lower than the U.S., U.K., and Europe in context prescience, suggesting an educational and research commitment to drawing ideas from the traditional disciplines (e.g., physics) or expected combinations (e.g., chemistry + material science), with India showing the most notable recent increase (see Figs. S4–S7 for robustness checks across databases, country exclusions, thresholds, and author attribution methods).

Where *emergence* captures predictable developments resulting from detectable currents and trends in techno-scientific discovery and invention, *prescience* captures unpredictable departures from those trends that presage new trends. In the context of strategic techno-scientific investments of attention and resources, the first is more assured, and the second is more risky. If those investments are successful, however, the second reshapes the technoscientific landscape, while the first predictably extends existing trends.

Figure 3A illustrates the techno-scientific attention landscape. New papers that interlink improbable concepts from the left and right peaks at time zero, become prescient insofar as those peaks approach one another and increase the predictability of future work following in that tradition at time one. 3B and 3C show the share of all prescient papers attributable to selected countries. Note that we calculate 2 different kinds of prescient recombination. The first, context prescience, measures the novelty of combinations of journals in a paper's reference list, representing the diversity of disciplines the work draws on. The second, content prescience, measures the novelty of combinations of author-supplied keywords that describe substantive aspects of the work. Note that a paper with authors from multiple countries is attributed to all of the authors' countries equally, so the shares do not sum to 100. Within these facets, inset plots trace the total number of prescient publications per country to remind readers that all countries are growing their research portfolios and that no country is in decline. Panels 3D and 3E show the per-paper rates of prescience for each country. This allows a more direct comparison of each country's research character, net of total publication volume.

## Overall Landscape of Scientific Advancement

Emergence, prescience, and disruption, combined with production rates, yield an integrated global landscape of scientific advancement. Trends in these measures parallel traditional metrics (Figure 4), but critically, the timing of shifts varies across measures. While Chinese papers grew rapidly in volume, their citation share lagged years behind. This is partly mechanical—China had less prior work to cite. But citations per paper reveal Chinese papers were at the bottom among major producers until 2010, reaching parity with the EU in 2016 and the US in 2021. This conforms with findings that frontier chemical research receives a citation discount (*22*).

A similar pattern emerges when comparing emergence with prescience. Emergence—observing trends and extending them predictably—is straightforward though not trivial. China moved from bottom to top in the rate of emerging work during 2010–2020, reflecting investment in leading-edge R&D, especially in critical technology areas (Fig. S8). China lags, however, in prescience: radical departures that precede new areas. National research cultures and funding incentives may encourage interdisciplinary work more in some countries (US, UK, EU, India) than others. Prestige may also matter: unprecedented work may more likely launch trends when it originates from high-status countries and institutions—which could explain US and UK leadership in prescience, though not India's.

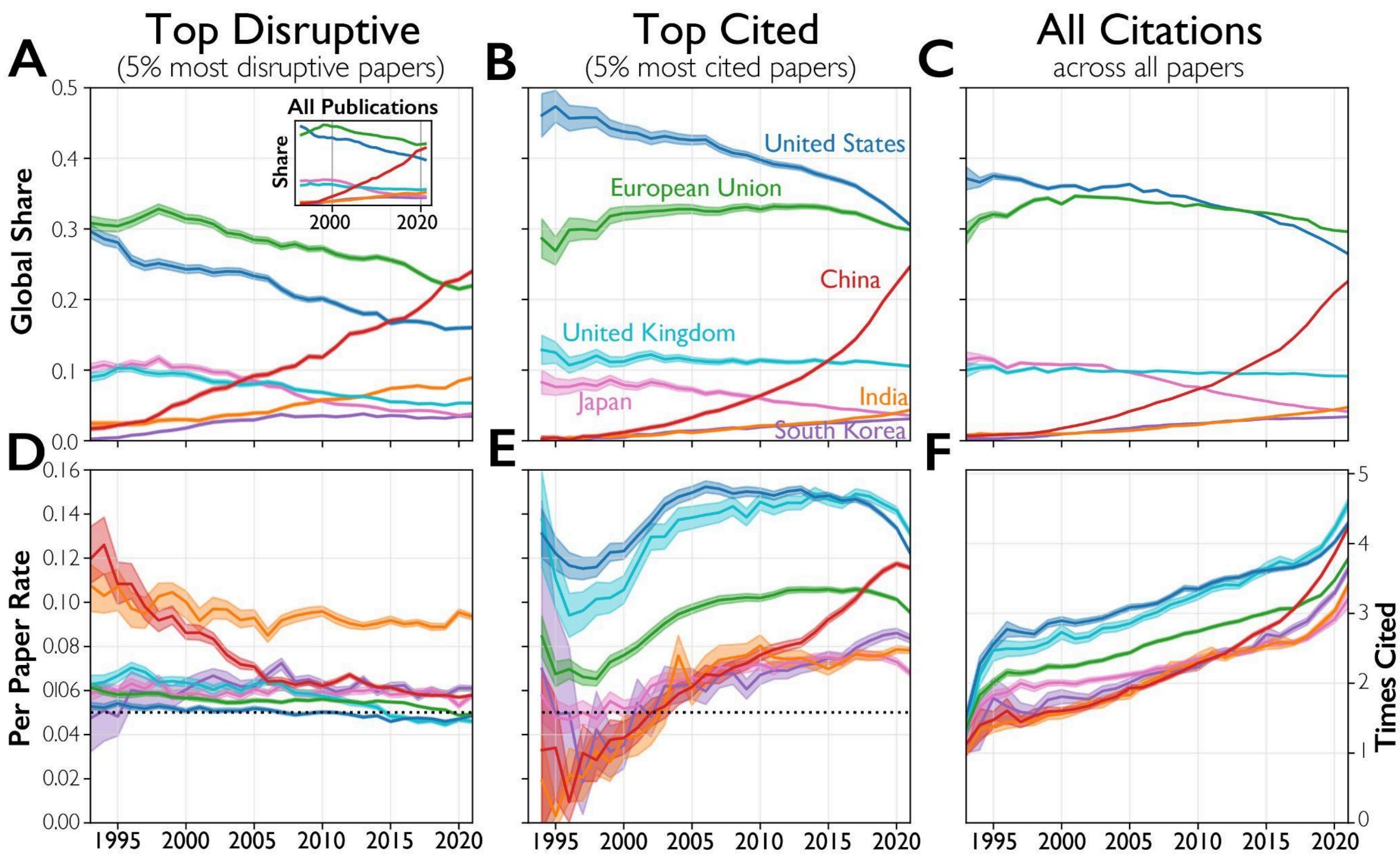

**Figure 4. Disruption, Citation, and Publication.** **A**-**C**, Global share of the top 5% most disruptive papers, the top 5% most cited papers, and of all citations across the world's papers, respectively, with an inset of the global share of publications. Total publications and share of the most disruptive papers track closely, with the U.S. and E.U. steadily declining relative to China's rise, leading the world in top-disruptive share by 2018 (**A**). Despite having a lower publication share than Europe, until very recently, the U.S. has a higher share of top and overall citations. China is rapidly approaching the U.S. and E.U. in top and overall citations (**B**-**C**). **D**-**F**, Per paper rate of top 5% disruptive papers, top 5% most cited papers, and overall citations. We see an inverse relationship between China's trend of dropping disruption and rising citations, which contrasts with India's sustained disruption share and rising citations. The U.S. and U.K. lead in per paper citations, but China surpassed Europe in top and overall cites per paper in 2017, and is nearly at parity with the U.S. and U.K.

# Discussion

We take two complementary approaches to evaluating national roles in scientific advance. First, we embed research pathways in a hypergraph of scientists and concepts to identify leadership in emerging areas. Second, we embed papers to capture the (im)probability of concept combinations, identifying prescient work. We compare these approaches with established metrics of volume, citation, and disruption indices across the citation graph to identify work recognized as catalyzing new directions versus developing established ones. We apply these across all areas and within strategic domains: AI, energy, biotechnology, and semiconductors (Fig. S8).

These complementary measures reveal that nations do not simply lead or lag in science—they lead *differently*. China's dominant strategy has been to identify and amplify nascent consensus: investing massively in areas where convergence is already detectable, and staffing those areas at scale. This is neither trivial nor merely imitative. Anticipating which emerging areas will prove consequential requires sophisticated judgment, and executing at speed demands institutional coordination that few systems can match. The result is that China shapes the trajectory of global science not by redirecting it, but by accelerating it—arriving at the frontier first by reading the landscape most effectively.

The United States and Europe pursue a complementary but distinct form of leadership. Their persistent advantage in prescient work—research that departs from predictable trajectories and prefigures future directions—reflects institutional cultures that reward disciplinary boundary-crossing, tolerate failure, and confer prestige on surprising results. This is particularly visible in context prescience, where drawing on unlikely combinations of intellectual traditions remains a pronounced Western strength. The gap is not explained by volume alone; it persists in per-paper rates, suggesting differences in research culture rather than mere scale.

These patterns carry direct implications for science policy and funding strategy. Nations seeking to lead emerging areas benefit from centralized horizon-scanning, rapid resource mobilization, and large coordinated teams—capacities at which China's system currently excels. Nations seeking to generate prescient advances require different institutional commitments: sustained support for curiosity-driven research, interdisciplinary training and hiring, funding mechanisms that tolerate high-risk work with uncertain timelines, and evaluation criteria that do not penalize surprising results. Vannevar Bush advocated many of the correlates associated with this second approach to U.S. President Franklin D. Roosavelt in his proposal for an exploratory National Science Foundation in 1945 (*10*). The two strategies are not mutually exclusive, but they do compete for attention and resources within national research systems, and optimizing for one can erode the conditions that sustain the other.

This tension is visible in the data. China's rising emergence share coincides with relatively flat or modest gains in prescience per paper, particularly in context prescience. Conversely, the United States and Europe have ceded ground in emerging areas even as they maintain prescient leadership, suggesting that the decentralized, investigator-driven model that generates unexpected breakthroughs may be less effective at rapidly concentrating effort when a promising direction becomes legible. In the language of statistical test theory, an emergence policy minimizes Type 1 error such that more of what it proposes will succeed, but an undirected prescience strategy will necessarily more effectively minimize Type 2 error in that more of the best possible directions will be proposed. Whether either system can cultivate both capacities simultaneously remains an open question with significant consequences for technological competition.

Finally, these findings caution against relying on any single metric of scientific leadership. Publication counts and citations reflect past investments. Disruption indices capture community reception but not prospective direction. Emergence and prescience offer forward-looking assessments, but they measure different things—one tracks the ability to read the scientific landscape, the other the ability to reshape it. A nation that dominates in volume and emerging areas but lags in prescient work is well-positioned to exploit known opportunities but will be poorly positioned to create new ones. The reverse may also be true. Comprehensive assessment of scientific competitiveness requires all of these measures, and the strategic choices they illuminate are ultimately choices about what kind of scientific future a nation will build.Methods

# Data

We use the Web of Science (WoS) database for our modeling. We focus most of our analyses on the years 1990-2023 as the author-generated keyword data, required for our surprise and prescience analyses, is sparse before 1990 (see Fig. S1). Author-based analyses begin in 2008, a year when WoS indexing and metadata linkage for authors dramatically improved, and end with the most recent year for which we have full data. We excluded review articles and non-English articles. We replicate these analyses with OpenAlex (*23*), an open access database similar to the WoS, but with broader coverage of publication venues, especially including conferences where leading AI research is published. Results of the OpenAlex replications are detailed in the Supplementary Information and broadly confirm the pattern of findings from WoS.

# Modeling Emergence

A scientific area is formed by taking the union of a central keyword and its 24 nearest neighbors in a trained embedding space, yielding 25 keywords per area. We define an area as emerging in a given year if it has the following properties: rapid growth in amount of collective attention that the area's central keyword receives over the preceding years (*frequency growth*) resulting in a substantially large count of mentions in the current year (*prevalence*), as well as an increasing trend in pairwise semantic similarities among the nearby keywords within a recent time frame (*keywords convergence*). The embedding space is constructed using the expert-aware word embedding approach (*15*) that runs the deepwalk algorithm (*24*) on the literature's hypergraph to obtain an embedding vector per keyword. The hypernodes in this network consist of authors and keywords, and hyperedges are publications such that a set of keywords and authors are connected if they belong to the same paper. Nearest neighbors of a given keyword were obtained in the resulting embedding space. The prevalence of each keyword in a given field is simply computed as the number of times authors used it in the year of analysis in the field's literature, normalized by the total count of all keywords in the field. The growth rate was obtained by fitting a temporal exponential function of form $ae^{bt} + c$ to the keyword's annual frequencies in the same field computed over the specified 5-year time window, with coefficient $b$ of the fitted function used as the frequency growth rate metric. Keyword convergence quantifies how quickly nearby keywords in the area move closer together in the embedding space, based on average pairwise distance velocities over the past three years. Faster convergence, measured by a substantial decrease in the average pairwise distances, implies that the area's keywords have been coming together to form the area more quickly. Note that while the prevalence and frequency growth of a certain keyword depend on the field we are focusing on, the convergence score is computed independently of the field, as the embedding space is computed universally, considering all available publications.

To label an area as emerging, we require high scores on all three criteria. Moreover, in order to rule out areas whose central keywords' mention frequency grows sub-exponentially, we quantified the quality of exponential fitting through $R^2$ and added it as the fourth metric. We computed four separate rankings by sorting the areas based on each of the four metrics. These individual rankings were then combined by averaging each area's positions to obtain a single score per area. Areas with higher final scores are likely to have a central keyword at the center of a coalescing set of keywords and to show substantial frequency growth and prevalence. We select the top 1% of areas with the highest final scores to define emerging fields in techno-scientific discovery and invention. To account for the size of different scientific disciplines, we ran the above analysis for each discipline independently. We considered 19 fields used by modern bibliographic databases, such as OpenAlex, as the main scientific disciplines.

In order to decide whether to mark a paper as emerging, we first computed every paper's distance to emerging areas. For this purpose, we selected only papers with at least one keyword and used the subject labels from the Web of Science to assign disciplines. We then considered all areas already marked as emerging within the same set of disciplines and computed the distance between the paper and these emerging fields. To quantify distances, we represented papers and areas using the centroids of their keyword embeddings and computed cosine distances. The ultimate distance between a given paper and a set of emerging areas was defined as the minimum distance between the paper's centroid embedding and those of other areas. Emerging papers were defined as the top 5-percentile of those with the smallest distance to emerging areas.

## Modeling Prescience

A new combination of research contents may surprise because it has never succeeded before, despite having been considered and attempted (*25–28*); or because it has never been imagined—representing a combination of ideas inaccessible within disciplinary conversation. Our model articulates a simple mental process: Scientists and inventors combine things together that are (1) scientifically or technologically close and (2) cognitively salient. While formally simple, this model is more effective than those from the literature and even advanced deep learning models in capturing novelty and impact (*16*). Specifically, we model the likelihood that keywords or sources become combined as a function of (1) their *proximity* in a latent embedding space and (2) their *salience* to scientists through prior usage frequency. Specifically, each node $i$ is associated with an embedding (or latent vector) $\theta_i$ that embeds the node in a latent space constructed to optimize the likelihood of observed papers and patents. Each entry $\theta_{id}$ of the latent vector denotes the probability that node $i$ belongs to latent dimension $d$. Dimensions underlying these latent vectors naturally recover scientific fields; see (*16*). The complementarity between contents in hyperedge $h$ (a paper) is modeled as the probability that content nodes (i.e., keywords) load on the same dimensions, $\sum_d \prod_{i \in h} \theta_{id}$. We measure these with both author-generated keywords, and keywords imputed by classification, finding similar results. The same is true for context models, where the contexts are journals or conferences cited as a source of ideas. We associate each node $i$ with a latent scalar $r_i$ to account for its cognitive salience or the exposure scientists have had to it, measuring its overall connectivity in the network. We then model the propensity ($\lambda_h$) of combination $h$—our expectation of its appearance in actual papers and patents—as the product of proximity between nodes in $h$ and their salience:

$$\lambda_h = \sum_d \prod_{i\in h} \theta_{id} \times \prod_{i\in h} r_i \tag{1}$$

Then the number of publications or patents that realize combination $h$ is modeled as a Poisson random variable with $\lambda_h$ as its mean. Finally, the likelihood of a hypergraph $G$ is the product of the likelihood of observing every possible combination.

Across biomedical sciences, physical sciences, and inventions, the model correctly distinguishes between a content combination that turned into a publication and a random combination more than 95% of the time based on data from previous years (Biomedicine: AUC=0.98; Physics: AUC=0.97; Inventions: AUC=0.95) (*16*). The successful classification of future combinations suggests that the embedding spaces we construct from frequent combinations of contents and sources in published research inscribes a space of latent knowledge. Researchers tend to wander locally across this space in generating new papers and patents. This agrees with previous findings regarding inertia and conservative search in science (*25*, *29–32*).

With a measure of what science and technology is expected, we assess the novelty of a combination $h$ as the improbability or surprisal (*33*) of its assemblage:

$$novelty(h) = -\log \sum_d \prod_{i\in h} \theta_{id}. \tag{2}$$

As we model contents and sources separately—in one set of models, a paper is a combination of contents (e.g., keywords) and in another it is a combination of sources (e.g., referenced journals)—we also separately measure the novelty of its content combination and its source combination.

We measure the prescience of a paper by the decrease in its surprisal between when it was published and two years later, in order to identify work that was initially surprising, but became the norm through leadership or luck. As the hypergraph embedding model described above is dynamic and the underlying embedding space varies by year, we compute two surprise scores for each paper: one using the embeddings from the year when it was published and the other using the embeddings from two years later. This allows us to measure prescience of the paper as the difference between the two surprise scores: papers that were surprising initially but become much less so later on are papers that were prescient or 'ahead of their time,' unusual at their publication date but typical of future work. As with emerging areas, we define a paper as prescient if it is in the top 5 percentile of papers in its field. In the SI, we show the results for other thresholds.

## Measuring Disruption

To capture how the innovativeness of papers are recognized by the scientific community, we calculate the disruption index (*18*, *19*, *21*) of scientific publications based on citation networks. The disruption index D, which characterizes to what extent the focal paper destabilizes or consolidates prior science, can be computed as

$$D = \frac{n_f - n_b}{n_f + n_b + n_r}$$

where $n_f$ is the number of subsequent papers citing only the focal paper and none of its references, $n_b$ is the number of subsequent papers citing both the focal paper and its references, and $n_r$ is the number of subsequent papers citing only the references of the focal paper. Varying between -1 and 1, the disruption index portrays the degree to which the focal paper disrupts or develops science. If a work completely overshadows prior work by garnering all subsequent attention to itself while its inspirations lie forgotten, then the measure equals 1. If a work is always cited alongside its inspirations and visa versa, then the work can be understood to broadcast the importance of prior work and the measure equals -1. For example, IBM's 1982 U.S. patent for a Scanning Tunneling Microscope and Cetus' 1987 U.S. patent for a "Process for Amplifying Nucleic Acid Sequences" or the Polymerase Chain Reaction (PCR), were both associated with Nobel Prizes, and both post a disruptiveness score of 1, while Amazon's successful *One-Click* patent for online purchasing is much less than one, and amplifies prior e-commerce innovations (*18*). A larger disruption index implies the focal paper is more likely to disrupt existing ideas and its innovativeness is more favorably appreciated by the scientific community; while a smaller one reveals the focal paper tends to develop or consolidate prior knowledge. To make the comparisons fair, we compute disruption scores only for papers with both references and citing papers. Similar to existing literature (*18*, *19*, *21*), we use the $CD_5$ index, i.e., the disruption index with 5-year time windows where only papers published within 5 years from the focal paper's publication are considered in the calculation of $n_f$, $n_b$, and $n_r$. We define disruptive papers as the ones that are in the top 5 percentile of their field, to mirror our other analyses.

## Figures

Figures show annual proportions. Where appropriate, 95% confidence intervals have been calculated and drawn as shaded regions around the lines.

# Data Availability

Web of Science data are available from Clarivate Analytics. OpenAlex data are publicly available here: [https://openalex.org](https://openalex.org)

# Acknowledgments

We acknowledge support from the National Science Foundation (grants 2404109 for J.E. and 2241237 for J.E. and J.L.), the Air Force Office of Scientific Research (FA9550-19-1-0354 and FA9550-15-1-0162 for J.E.), and the DARPA grant "Modeling and Measuring Scientific Creativity" (J.E.). We thank members of the National Network for Critical Technology Analysis and the policy members with whom it engaged for motivation and feedback.

# Contributions

J.L., J.S., and J.E. jointly conceived and designed the study, interpreted the results, and drafted and revised the manuscript. F.S. provided conceptual, modeling and software support. J.L. organized the data, and J.L. and J.S. performed the data analysis. J.L. and J.E. designed and constructed the figures.

# References

1. P. A. David, D. Foray, Economic fundamentals of the knowledge society. *Pol. Futur. Educ.* **1**, 20–49 (2003).

2. J. E. Stiglitz, "Knowledge as a Global Public Good" in *Global Public Goods* (Oxford University PressNew York, 1999), pp. 308–325.

3. J. Mokyr, *Gifts of Athena* (Princeton University Press, Princeton, NJ, 2011).

4. J. A. Schumpeter, Capitalism, socialism and democracy (1942). *J. Econ. Lit.* **20**, 1463 (1976).

5. U. Akcigit, J. Van Reenen, *The Economics of Creative Destruction: New Research on Themes from Aghion and Howitt* (Harvard University Press, 2023).

6. W. Sombart, Krieg und kapitalismus (War and capitalism). *Munich and Leipzig: Duncker & Humblot* (1913).

7. R. R. Nelson, N. Rosenberg, Technical innovation and national systems. *National innovation systems: A comparative analysis* **322** (1993).

8. C. Freeman, The "National System of Innovation"in historical perspective. *Cambridge J. Econ.* **19**, 5–24 (1995).

9. OECD, *Manual de Frascati 2015* (OECD, 2018).

10. V. Bush, Science the endless frontier: a report to the President by Vannevar Bush. *Director of the Office of Scientific Research and Development, United States Government Printing Office, Washington* (1945).

11. P. Kennedy, *The Rise and Fall of the Great Powers* (Vintage Books, New York, 1987).

12. D. C. Mowery, N. Rosenberg, *Technology and the Pursuit of Economic Growth* (Cambridge University Press, Cambridge, England, 1989).

13. F. Shi, J. G. Foster, J. A. Evans, Weaving the fabric of science: Dynamic network models of science's unfolding structure. *Soc. Networks* **43**, 73–85 (2015).

14. J. Sourati, J. Evans, Accelerating science with human versus alien artificial intelligences, *arXiv*

*[cs.AI]* (2021). http://arxiv.org/abs/2104.05188.

15. J. Sourati, J. Evans, Accelerating science with human-aware artificial intelligence, *arXiv [cs.AI]* (2023). http://arxiv.org/abs/2306.01495.

16. F. Shi, J. Evans, Surprising combinations of research contents and contexts are related to impact and emerge with scientific outsiders from distant disciplines. *Nat. Commun.* **14** (2023).

17. B. Hofstra, V. V. Kulkarni, S. Munoz-Najar Galvez, B. He, D. Jurafsky, D. A. McFarland, The Diversity–Innovation Paradox in Science. *Proceedings of the National Academy of Sciences* **117**, 9284–9291 (2020).

18. R. J. Funk, J. Owen-Smith, A Dynamic Network Measure of Technological Change. *Manage. Sci.* **63**, 791–817 (2016).

19. L. Wu, D. Wang, J. A. Evans, Large teams develop and small teams disrupt science and technology. *Nature* **566**, 378–382 (2019).

20. Y. Lin, J. A. Evans, L. Wu, New directions in science emerge from disconnection and discord. *J. Informetr.* **16**, 101234 (2022).

21. M. Park, E. Leahey, R. J. Funk, Papers and patents are becoming less disruptive over time. *Nature* **613**, 138–144 (2023).

22. S. Qiu, C. Steinwender, P. Azoulay, Who Stands on the Shoulders of Chinese (Scientific) Giants? Evidence from Chemistry (2022). https://doi.org/10.3386/w30772.

23. J. Priem, H. Piwowar, R. Orr, OpenAlex: A fully-open index of scholarly works, authors, venues, institutions, and concepts. arXiv [Preprint] (2022). https://doi.org/10.48550/arXiv.2205.01833.

24. B. Perozzi, R. Al-Rfou, S. Skiena, "DeepWalk: online learning of social representations" in *Proceedings of the 20th ACM SIGKDD International Conference on Knowledge Discovery and Data Mining* (Association for Computing Machinery, New York, NY, USA, 2014)*KDD '14*, pp. 701–710.

25. A. Rzhetsky, J. G. Foster, I. T. Foster, J. A. Evans, Choosing experiments to accelerate collective discovery. *Proc. Natl. Acad. Sci. U. S. A.* **112**, 14569–14574 (2015).

26. J. G. Foster, A. Rzhetsky, J. A. Evans, Tradition and Innovation in Scientists' Research Strategies. [Preprint] (2015). https://doi.org/10.1177/0003122415601618.

27. H. Rheingold, *Tools for Thought: The People and Ideas behind the Next Computer Revolution* (Simon & Schuster Trade, 1985).

28. D. Simandan, Being surprised and surprising ourselves: A geography of personal and social change. *Prog. Hum. Geogr.* **44**, 99–118 (2020).

29. M. Callon, A. Rip, J. Law, *Mapping the Dynamics of Science and Technology: Sociology of Science in the Real World* (Springer, 1986).

30. R. Guimerà, A. Díaz-Guilera, F. Vega-Redondo, A. Cabrales, A. Arenas, Optimal Network Topologies for Local Search with Congestion. [Preprint] (2002). https://doi.org/10.1103/physrevlett.89.248701.

31. T. E. Stuart, J. M. Podolny, Local search and the evolution of technological capabilities. *Strategic Manage. J.* **17**, 21–38 (2007).

32. T. Jia, D. Wang, B. K. Szymanski, Quantifying patterns of research-interest evolution. *Nature Human Behaviour* **1**, 1–7 (2017).

33. T. M. Cover, J. A. Thomas, *Elements of Information Theory* (John Wiley & Sons, 2012).

# Supplementary Information

## Materials and Methods

### 1.1 Web of Science Data

We use a copy of the Web of Science (WoS) core collection data from Clarivate Analytics for our main analyses. The analyses reported here focus on the years 2000-2023. We restricted our analyses to exclude research and review articles not in English.

Countries are assigned to papers based on the authors' institutional affiliations. If a paper has authors from multiple countries, we assign full credit to each country in the main analysis. For papers indexed before 2008, WoS provides a list of authors and a list of their affiliations, but does not link them. After 2008, it links affiliations to specific authors, allowing us to evaluate the effect of assigning countries to papers based on the location of the first, last, or corresponding author. The results are qualitatively similar.

### 1.2 Exclusion of non-English Sources

Any international analysis of scientific literature will be biased if it is limited to only one language. China, in particular, has a robust academic publishing industry in Chinese that is not captured in our analysis. Our restriction to English is partly a technical one: there simply are not sources of data that have comprehensive, multilingual indexing of academic publications, and even if there were, there would be challenges to linking substantive topics and areas across languages. However, the large majority of academic publishing worldwide is now in English (80% - 95%, depending on field), and that share is growing over time as English becomes increasingly dominant and incentivised in academic writing.(*1*, *2*) Moreover, studying China's rise to prominence within the English scientific literature, as we do here, is substantively interesting in its own right, even though it is not a complete representation of all research conducted in China.

### 1.3 Emerging areas

#### 1.3.1 Hypergraph Deepwalk

Deepwalk algorithm (*3*) consists of running a random walk sampling on a network followed by training a word embedding model on the resulting node sequences to learn a vectorial representation of the nodes. This representation-learning process is designed for regular graphs, in which each edge connects a pair of nodes. Hypergraphs differ structurally in that (hyper)edges connect multiple (hyper)nodes. Here, following the expert-aware embedding approach (4), we extended the DeepWalk algorithm to hypergraphs by running a generalized random walk, which differs from a standard random walk in that it includes an additional hyperedge sampling step before randomly selecting the next neighboring node. We generated 20 nodes per random walk sequence and balanced the number of author to non-author nodes by setting the non-uniform sampling parameter $\alpha$ in the expert-aware deepwalk algorithm equal to one. For each year of analysis, we constructed a hypergraph of the literature produced over the past 5 years. The number of random walk sequences generated for each year was set to the number of keywords involved in the literature within this time window. Figure S1 shows the growth in scientific vocabulary over time: keywords per year grew from under 250,000 in 1990 to over 750,000 by 2020, while the 5-year hypergraph windows—which smooth year-to-year variation—exceeded 1.5 million unique keywords.

### 1.3.2 Emergence Scoring of Keywords

Frequency-based scores, i.e., frequency growth and prevalence, are computed based on the year-wise count of mentions of the keywords in abstracts. The convergence score is computed differently and depends on the trained embeddings of the keywords. For any given keyword w with embedding $E_{\mathrm{w,t}}$ in year $t$, we considered its corresponding area, i.e., the nearest 25 keywords to w in year $t$. We then examined how pairwise distances between all keywords in the area have changed in the last three years. Specifically, for all pairs ($\mathrm{w_i}$, $\mathrm{w_j}$) such that $\mathrm{w_i}$ and $\mathrm{w_j}$ belong to the area corresponding to the central keyword w, we'll have:

$$\Delta_t(\mathrm{w_i}, \mathrm{w_j}) = \mathrm{d}(E_{\mathrm{wi},t}, E_{\mathrm{wj},t}) - \mathrm{d}(E_{\mathrm{wi},t-1}, E_{\mathrm{wj},t-1})$$

$$\Delta_{t-1}(\mathrm{w_i}, \mathrm{w_j}) = \mathrm{d}(E_{\mathrm{wi},t-1}, E_{\mathrm{wj},t-1}) - \mathrm{d}(E_{\mathrm{wi},t-2}, E_{\mathrm{wj},t-2})$$

$$\Delta_{t-2}(\mathrm{w_i}, \mathrm{w_j}) = \mathrm{d}(E_{\mathrm{wi},t-2}, E_{\mathrm{wj},t-2}) - \mathrm{d}(E_{\mathrm{wi},t-3}, E_{\mathrm{wj},t-3})$$

where d is a distance metric (cosine distance in our case). For an emerging area, we expect a contracting set of embeddings or, equivalently, negative difference in distances $\Delta$'s, since the pairwise distances are decreasing in year. In other words, $\mathrm{w_i}$ and $\mathrm{w_j}$ would become semantically closer to each other in year $t$ than in year $t$–1 and, therefore, $\mathrm{d}(E_{\mathrm{wi},t}, E_{\mathrm{wj},t}) \leq \mathrm{d}(E_{\mathrm{wi},t-1}, E_{\mathrm{wj},t-1})$. The more negative the distance differences $\Delta$'s, the more convergent the corresponding area. Hence, we computed the convergence score by taking the average of $\Delta$'s defined above across the last three years and across all the area's keywords.

Figure S2A shows that medicine had the largest number of emerging areas until the early 2000s, when engineering surpassed it and grew rapidly. We ran our emergence analysis separately for each topic, so a keyword could define an emerging area in one field while being absent or non-emerging in another. Overlapping emerging areas across topics were smaller in early years and grew over time. Figures S2B–C show pairwise overlap between fields for 2010 and 2020: the largest overlap shifted from engineering-chemistry (2010) to engineering-computer science (2020), reflecting the increasing centrality of computational methods. While the largest overlap in 2010 occurred for "engineering" and "chemistry", the largest number of shared emerging areas were for "engineering" and "computer science" in 2020.

## 1.4 Prescience Modeling

### 1.4.1 Papers in Declining Areas

While prescience involves papers in areas that become much more probable after they are published, we can also examine the opposite: papers that become less probable, or those in declining areas. We see a similar pattern as with prescient papers: Chinese scholars are overrepresented, while American and European ones are underrepresented. In other words, Chinese research has higher variance.

### 1.4.2 The Meaning of Prescience

Our prescience model identifies papers that were "ahead of their time" in that they brought together combinations of keywords or sources that were unexpected at the moment the focal paper was published but subsequently became routine. Importantly, we use a time window of 2 years after a paper is published. We use this relatively short time window out of necessity, because our intention is to identify the current state of scientific leadership; if we used a longer time window, we would only be able to identify historical prescient papers. The length of our time window affects what our prescience scores measure, and what they do not measure. In particular, our prescience scores identify papers that were *2* years ahead of their time.

### 1.4.3 Prescience vs Surprise

Prescience and surprise capture distinct aspects of novelty. Surprise measures how unusual a paper's combinations are at the time of publication; prescience measures whether those combinations subsequently become routine. Figure S3 shows the relationship between these measures and citation impact. For both content (keyword) and context (reference) measures, prescience correlates more strongly with the likelihood of appearing in the top 10% most-cited papers than does surprise. This pattern is especially pronounced at the highest percentiles: papers above the 95th percentile in prescience (our threshold for "prescient" papers in the main analysis, indicated by vertical dashed lines) are substantially more likely to be highly cited than papers that are merely surprising. This suggests that scientific impact comes not from being unusual per se, but from anticipating directions the field will take.

## 1.5 Design robustness

### 1.5.1 Effect of database (Web of Science vs OpenAlex)

For robustness, we repeat all of our WoS analyses on a different data set, OpenAlex (Priem, et al. 2022). OpenAlex is larger overall, with fewer restrictions on both the quality and type of publication venue than WoS, leading to more coverage of dissertations, reports, grants, and conference proceedings. Notably, OA does not have author-chosen keyword data, so instead of keywords, we use OA's automatic classification of publications into 65 thousand "concepts," many of which resemble traditional author keywords.

Our results with this data are qualitatively similar. Below we show prescience rates by country using fine-grained concepts (Level 3 and below, roughly equivalent to the level of detail in Web of Science author keywords).

Figure S4 compares results across Web of Science (top row) and OpenAlex (bottom row) for all four innovation measures. The qualitative patterns are consistent: China's share rises over time across all measures, surpassing the US and EU in emerging areas and content prescience, while remaining lower in context prescience and disruptive papers.

Results with OpenAlex concepts are qualitatively similar but somewhat different from those with keywords. Authors writing in their respective subfields have different perspectives and motivations than scientometric databases that classify all published work. Second, keywords are selected prior to publication, whereas concepts are applied retrospectively, with the benefit of hindsight. Our prescience modeling specifically aims to capture innovations that are not apparent within a field at the time but become evident in hindsight. Therefore, using a variable such as concepts, which is based on hindsight, may undermine the estimation. Finally, OpenAlex indexes a broader range of publications than Web of Science, including lower-status journals and other types of publications (dissertations, books, white papers, etc.). Differences in the sampling frame may also account for modest discrepancies between the two databases.

### 1.5.2 Effect of excluding domestic papers

The main results presented are based on a model of all papers. However, major scientific producers such as the United States, the European Union, and China account for a substantial share of the global scientific literature, so innovations and trends in one country may not spread to the rest of the world. We test this by fitting additional models, this time training them only on papers that do not have authors from the respective location, and then evaluating how papers would appear from the perspective of the outside world.

Figure S5 shows results when papers from each major producer (US, China, EU) are excluded from model training. Solid lines show full-sample results; faded lines show results with exclusions. Emerging areas and context-prescience measures remain largely unchanged regardless of which country is excluded, suggesting that these forms of innovation diffuse internationally. Content prescience, however, shows national specificity: each country appears less prescient in language trends from the perspective of the outside world (faded lines fall below solid lines). This indicates meaningful national cultures in the ways that new scientific work is described.

### 1.5.3 Effect of changing thresholds for 'top' papers

Our main results define innovative work as the top 5% of papers by each measure, but this threshold is arbitrary. Figure S6 shows results for three thresholds: top 1% (99th percentile), top 5% (95th percentile), and top 10% (90th percentile). Results for emerging areas and content prescience are stable across thresholds. For disruptive papers, US and EU trends are robust, but China's share is lower at higher thresholds, suggesting that Chinese disruptive papers are somewhat more concentrated in the 90th–95th percentile than the top 1%. Context prescience shows the opposite pattern: Chinese papers are concentrated in the 90th–95th percentile, while US and EU papers appear disproportionately in the 99th percentile.

### 1.5.4 Effect of author attribution

Assigning papers to countries is complex: papers often have multiple authors from different countries, and some authors hold affiliations in more than one country. Figure S7 shows results using five attribution strategies: corresponding author only, first author only, last author only, any author (full credit to all countries, as in our main analysis), and unanimous (only papers where all authors share the same country). The qualitative pattern of China's rising share in emerging areas and declining Western dominance remains consistent across all attribution methods, indicating that the results are not artifacts of international collaboration patterns.

## 1.6 Topic Areas

Figure S8 presents results for research domains identified as strategically important by the National Science Foundation and U.S. Congress, including biomedical technology/genomics/synthetic biology, AI/ML/autonomy, advanced communications, disaster prevention, data management/security, and energy. Top rows show shares of innovative papers; bottom rows show innovation rates relative to publication volume (with the 5% baseline indicated by dashed lines). Results reveal domain-specific variation: China leads in emerging areas across most domains, while the US maintains advantages in prescience measures, particularly in AI/ML and biomedical fields. Confidence intervals (shaded regions) are wider in smaller fields.

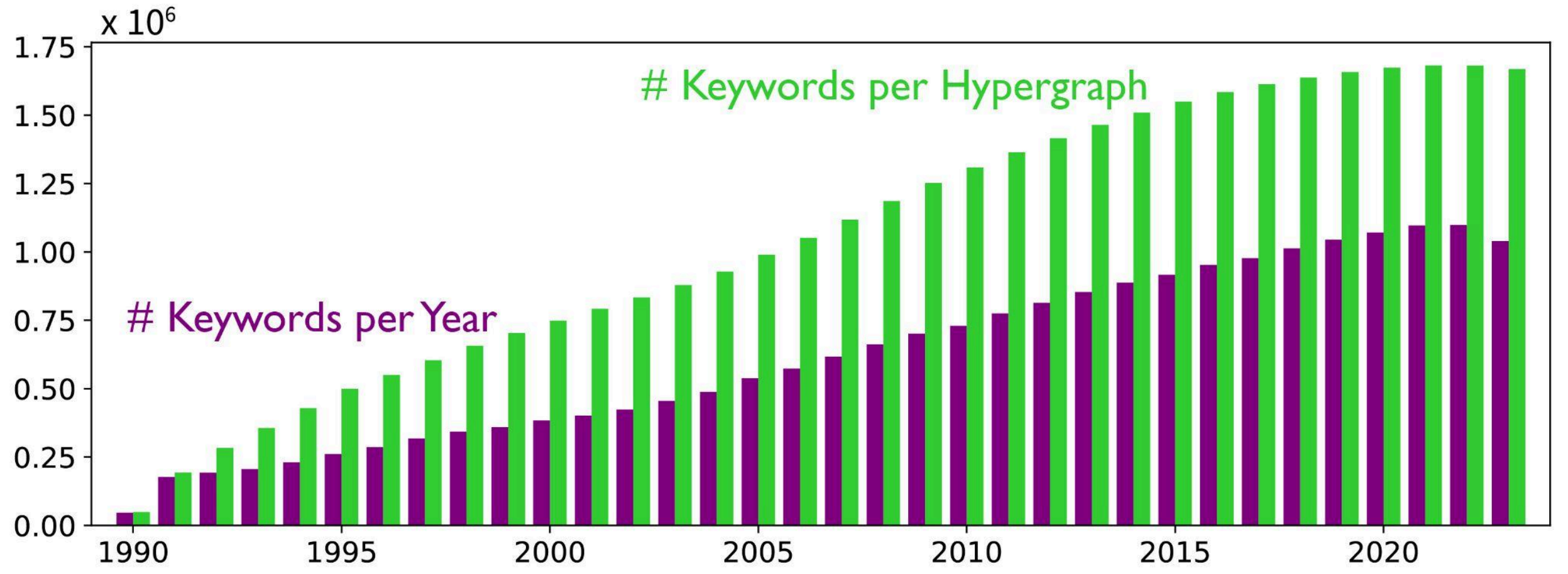

**Figure S1. Growth in scientific vocabulary captured by hypergraph embedding analysis.** Number of unique keywords appearing in the scientific literature by individual year (purple bars) and in hypergraphs constructed from rolling 5-year publication windows (green bars), 1990–2023. The 5-year hypergraph windows smooth year-to-year variation while capturing sufficient temporal context for embedding convergence analysis. Total vocabulary exceeded 1.5 million keywords by 2020, reflecting the rapid expansion and specialization of scientific research.

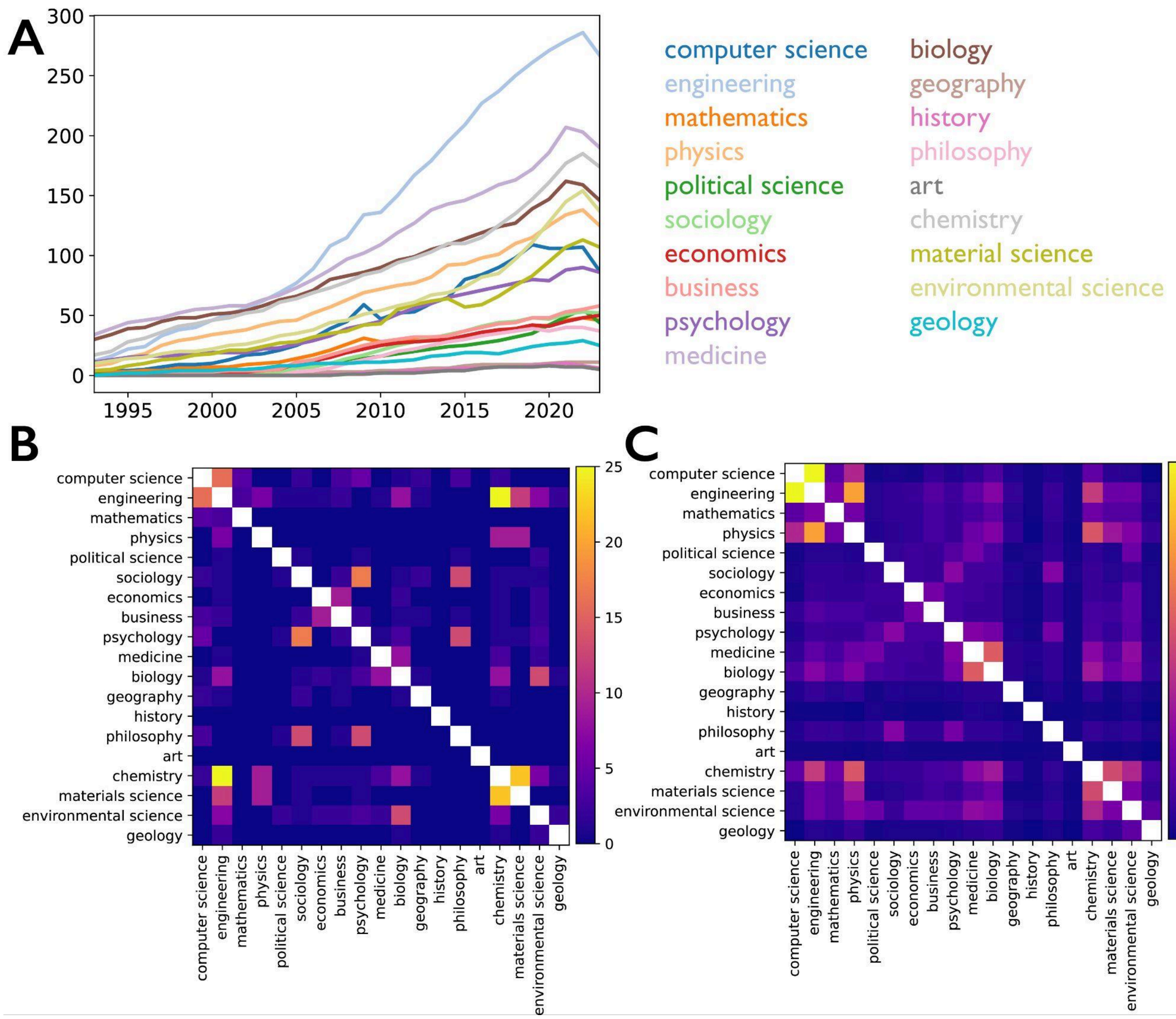

**Figure S2. Emerging research areas across scientific disciplines.** **A**, Number of emerging areas identified per year across 18 scientific subfields, 1993–2025. Medicine (dark blue) dominated emerging area production until the early 2000s, when engineering (light green) surpassed it and grew rapidly. Computer science (purple), including machine learning and generative artificial intelligence shows accelerating growth in recent years (see Fig. 2). **B**–**C**, Heatmaps showing the number of overlapping emerging keywords between pairs of disciplines in 2010 (**B**) and 2020 (**C**), with lighter colors indicating greater overlap. The largest interdisciplinary overlap shifted from engineering-chemistry in 2010 to engineering-computer science in 2020, reflecting the increasing centrality of computational methods across fields.

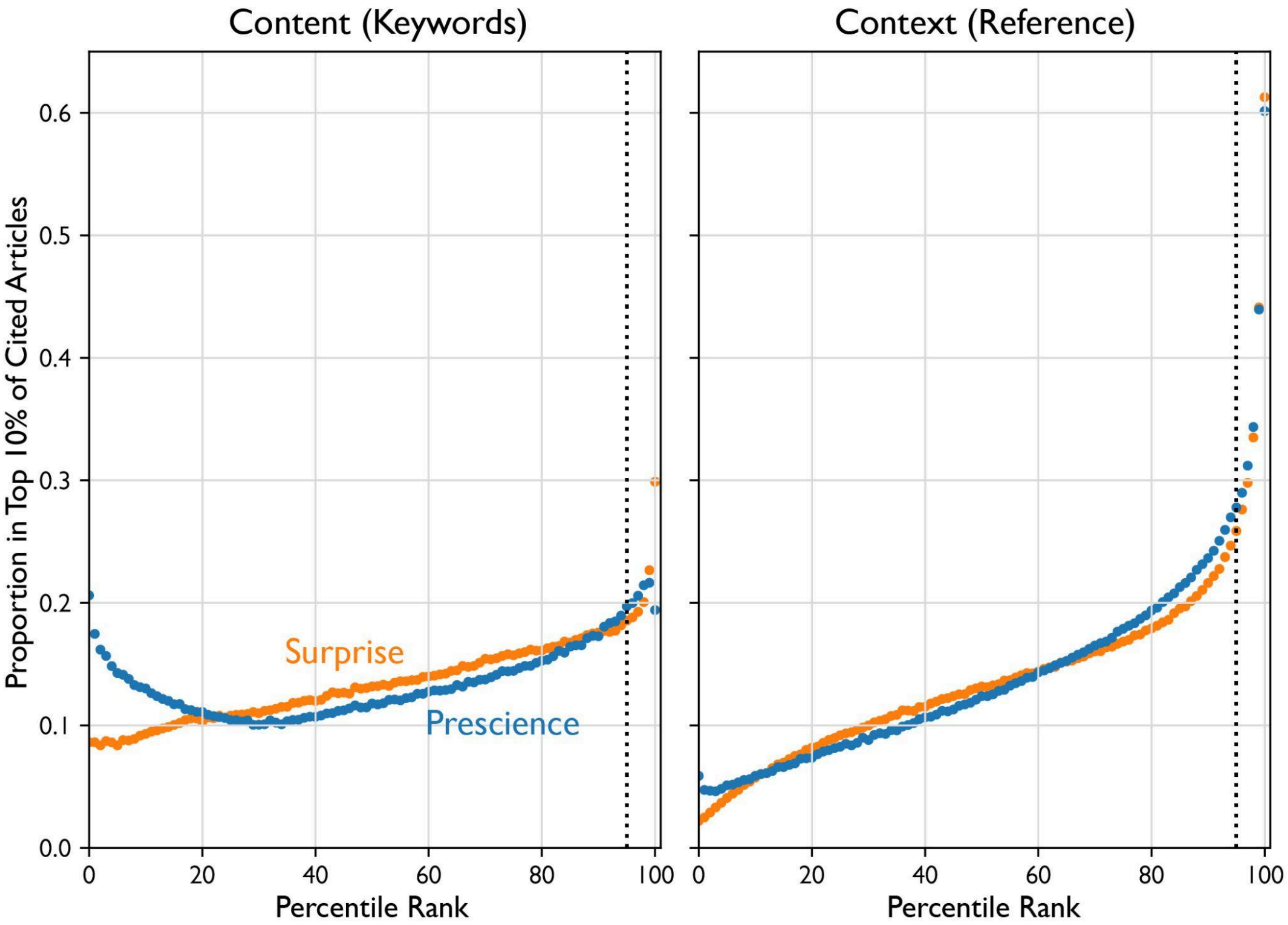

**Figure S3. Relationship between novelty, prescience, and citation impact.** Proportion of papers in the top 10% by citation count as a function of percentile rank for surprise (orange) and prescience (blue) measures. Left panel: content (keyword) metrics; right panel: context (reference) metrics. Vertical dashed lines indicate the 95th percentile threshold used to define prescient papers in the main analysis. Prescience posts a stronger positive relationship with citation impact than surprise, particularly at the highest percentiles, indicating that papers combining elements that subsequently become routine are more influential than papers that are merely unusual at the time of publication.

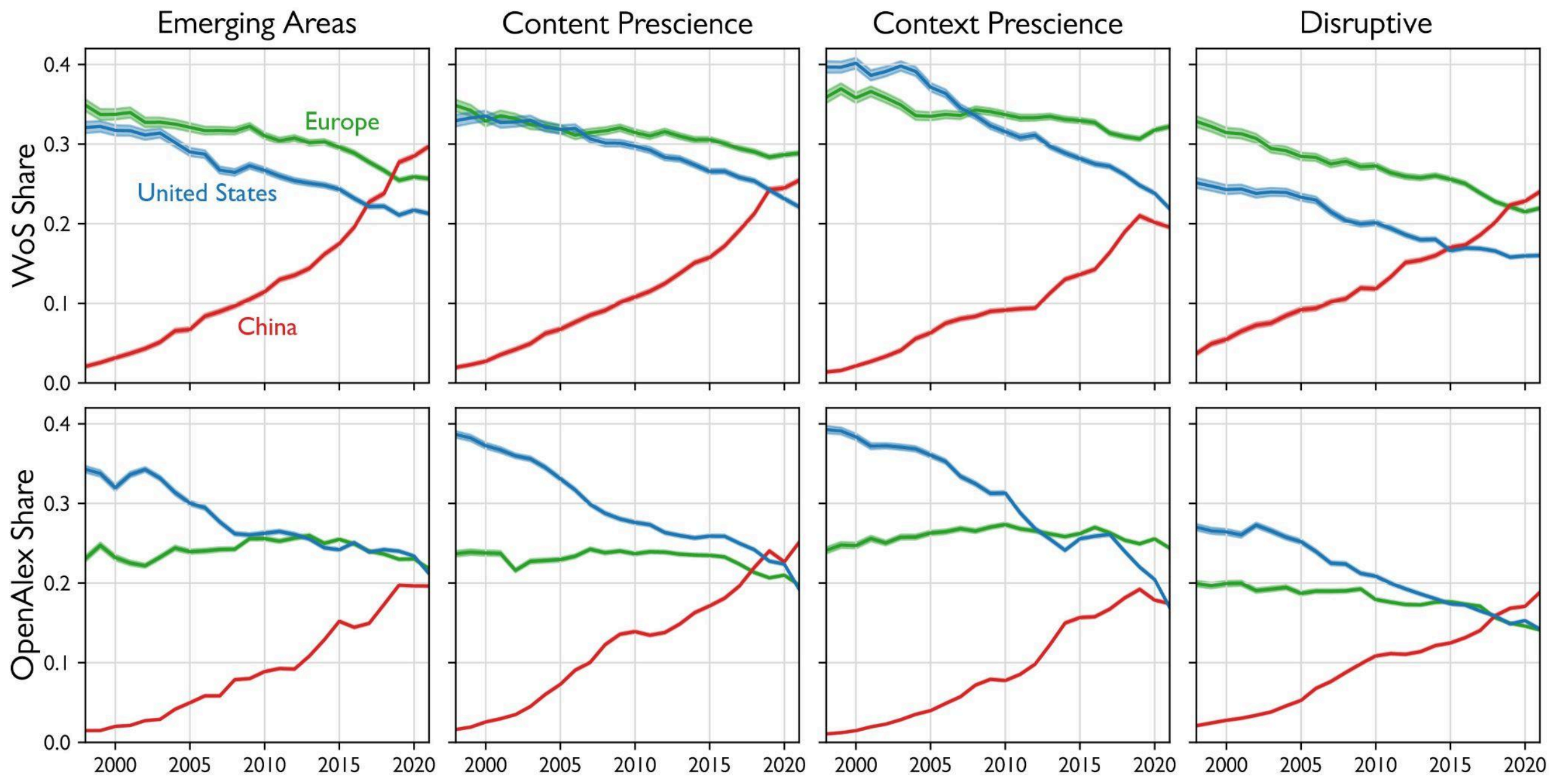

**Figure S4. Robustness of results across bibliometric databases.** Share of innovative papers by country (China in red, United States in blue, Europe in green) across four innovation measures: emerging areas, content prescience, context prescience, and disruptive papers. Top row: Web of Science (WoS) data. Bottom row: OpenAlex data. Despite differences in coverage—OpenAlex includes dissertations, reports, and conference proceedings excluded from WoS—the qualitative patterns remain consistent across databases. China's share rises across all measures over time, surpassing the US and the EU in emerging areas, content prescience, and disruption, while remaining lowest in context prescience.

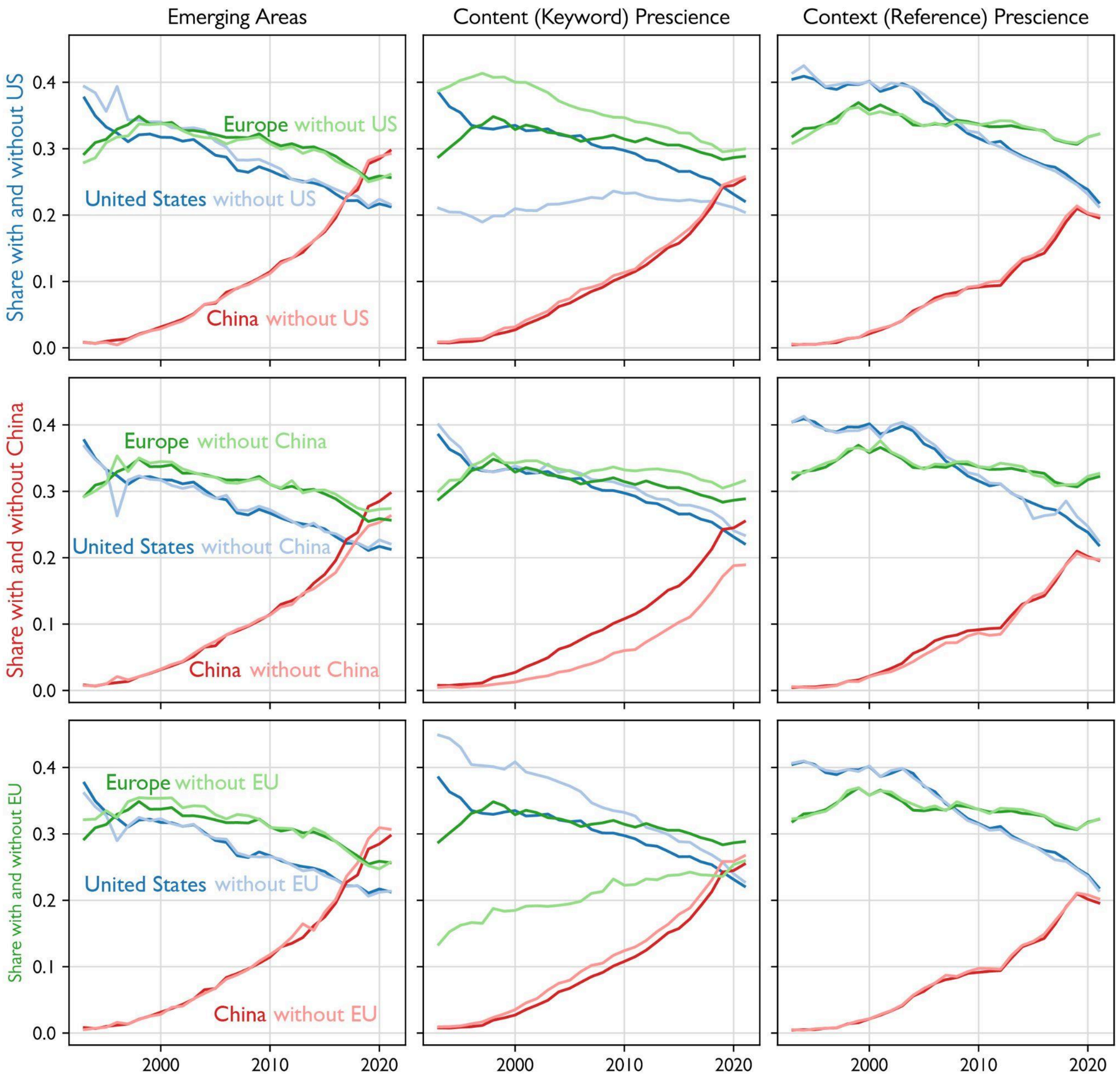

**Figure S5. Stability of innovation measures when excluding major scientific producers.** Each row shows results when papers from one major producer are excluded from model training: United States (top), China (middle), and European Union (bottom). Solid lines show full-sample results; faded lines show results excluding the indicated region. Emerging areas and context prescience measures remain largely unchanged regardless of which country or region is excluded, suggesting these forms of innovation diffuse internationally. Content prescience, however, shows moderate national specificity, such that each country appears less prescient in language trends from the perspective of the outside world, indicating meaningful national cultures in how scientific work is described.

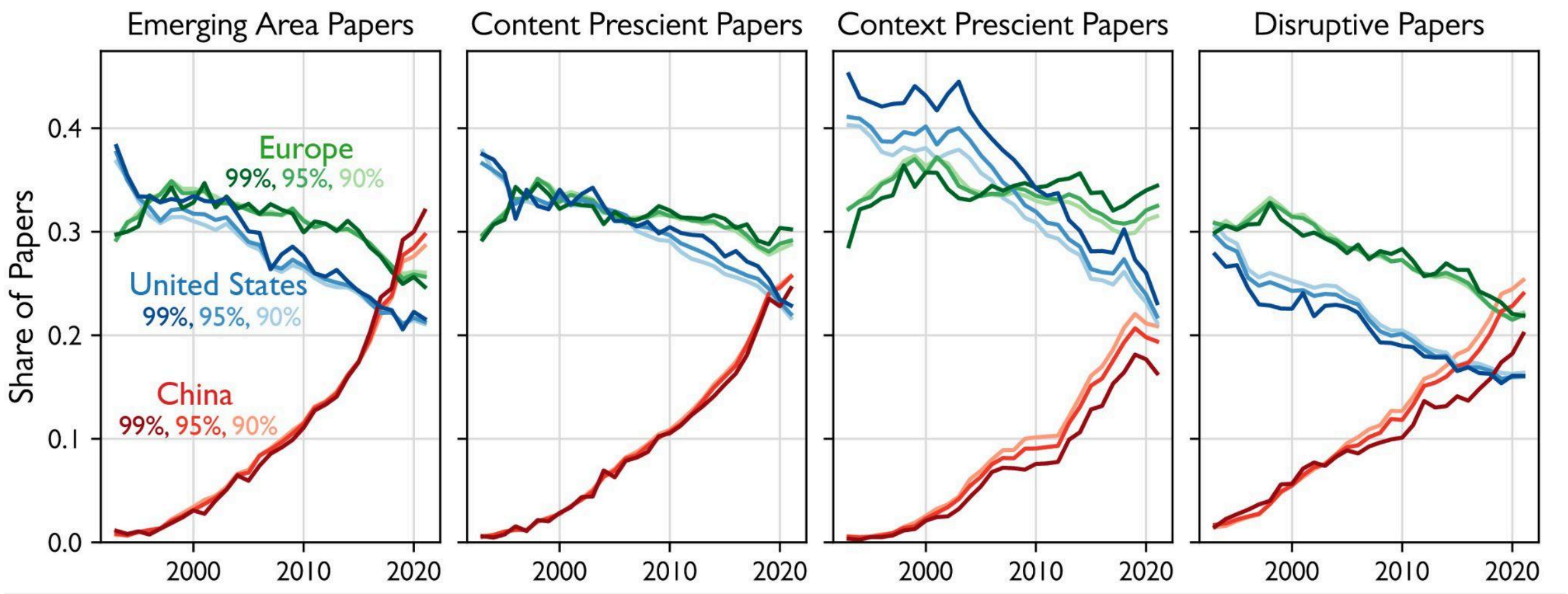

**Figure S6. Sensitivity of results to percentile thresholds for defining innovative work.** Share of papers by country across four innovation measures using three thresholds: top 1% (99th percentile, darkest), top 5% (95th percentile, medium), and top 10% (90th percentile, lightest). Results for emerging areas and content prescience are stable across thresholds. For disruptive papers, US and EU trends are robust, while China's share is lower at higher thresholds, suggesting Chinese papers are more concentrated in the 90th–95th percentile range. Context prescience shows the opposite sensitivity: Chinese papers are concentrated in the 90th–95th percentile, while US and EU papers appear disproportionately in the 99th percentile.

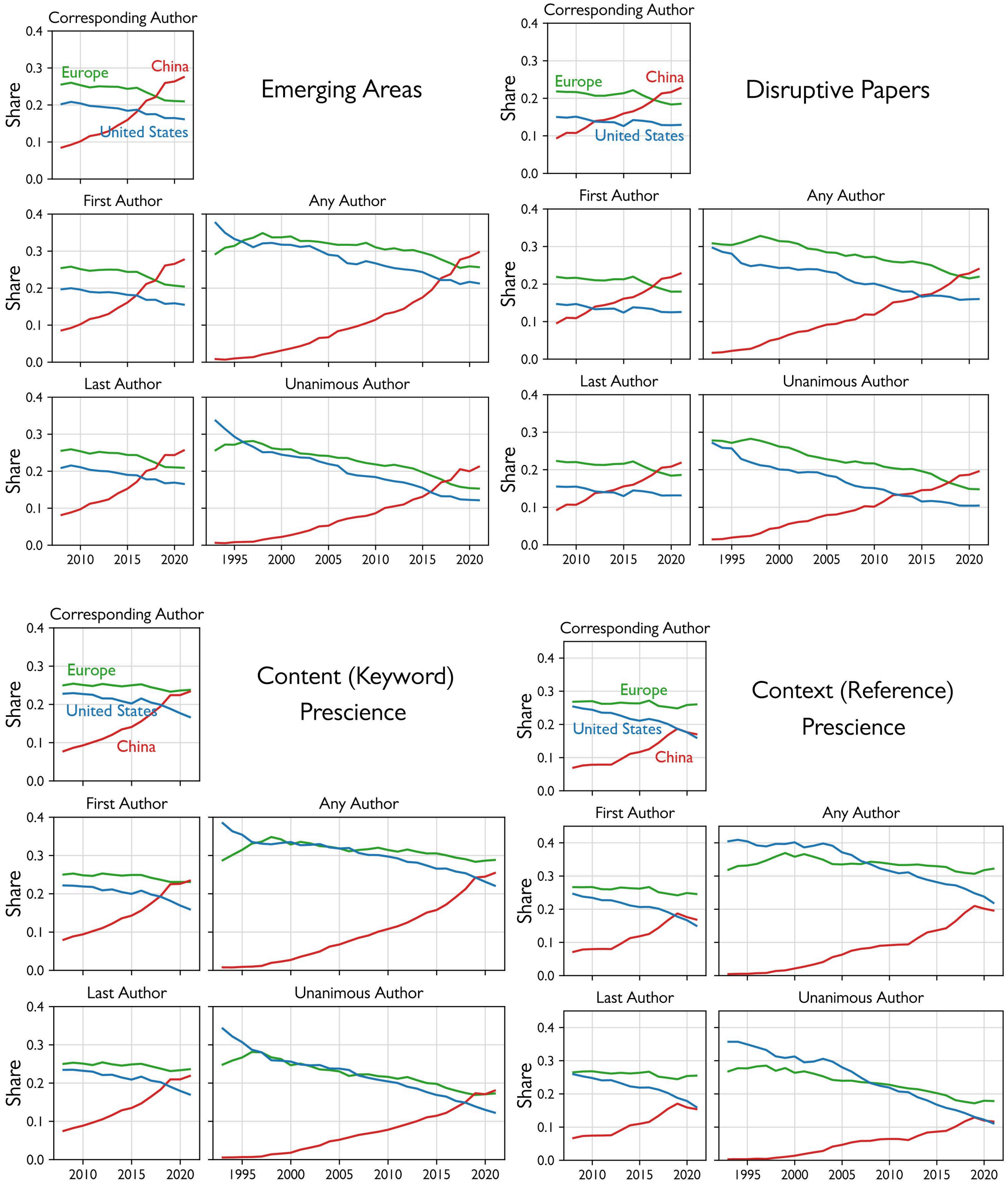

**Figure S7. Robustness of results to author attribution methods.** Results for four innovation measures (emerging areas, disruptive papers, context prescience, content prescience) using different strategies for assigning papers to countries: corresponding author only, first author only, last author only, any author (full credit to all countries represented), and unanimous (only papers where all authors are from the same country). The qualitative pattern of China's rising share in emerging areas and declining Western dominance remains consistent across all attribution methods, indicating that results are not artifacts of international collaboration patterns or authorship conventions.

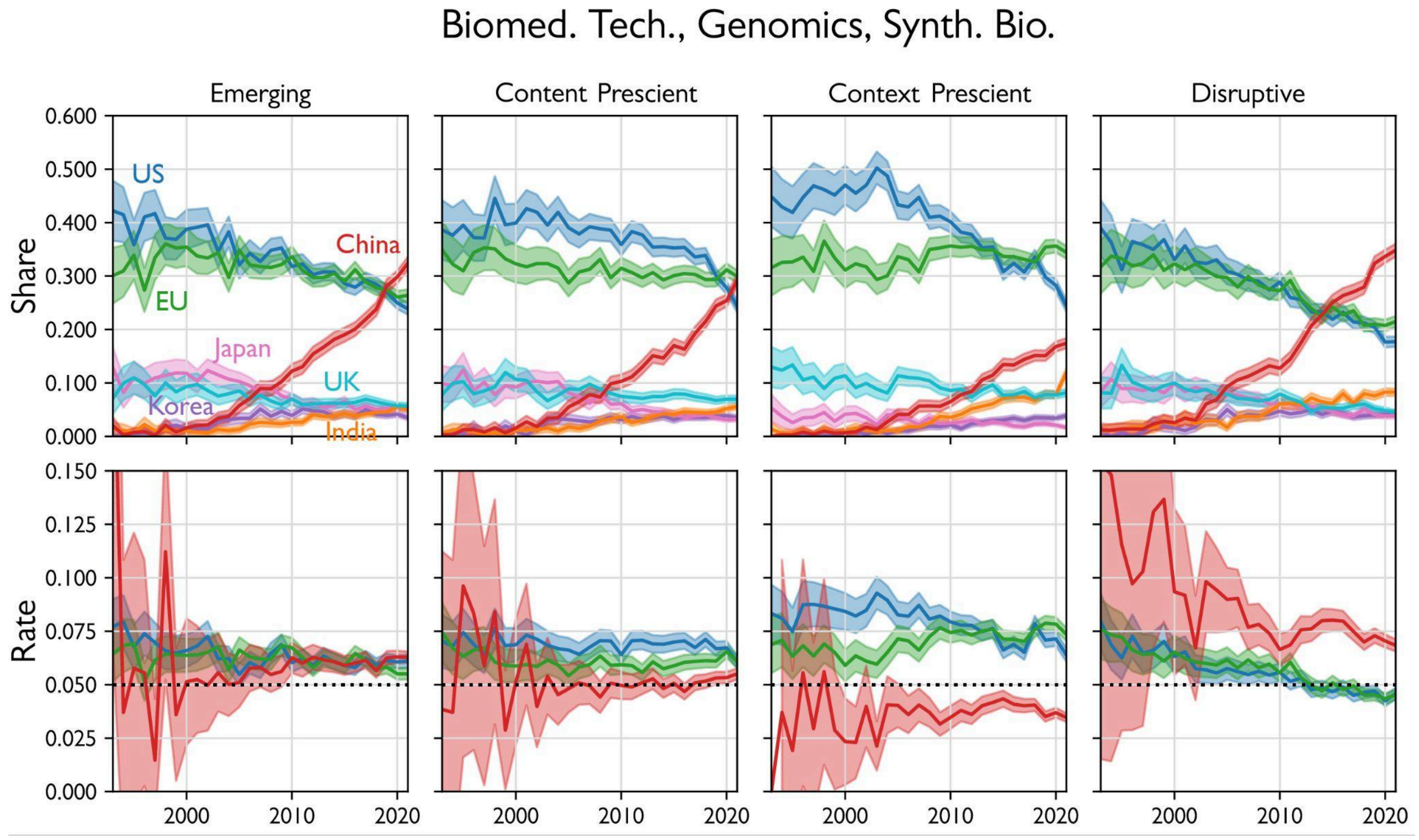

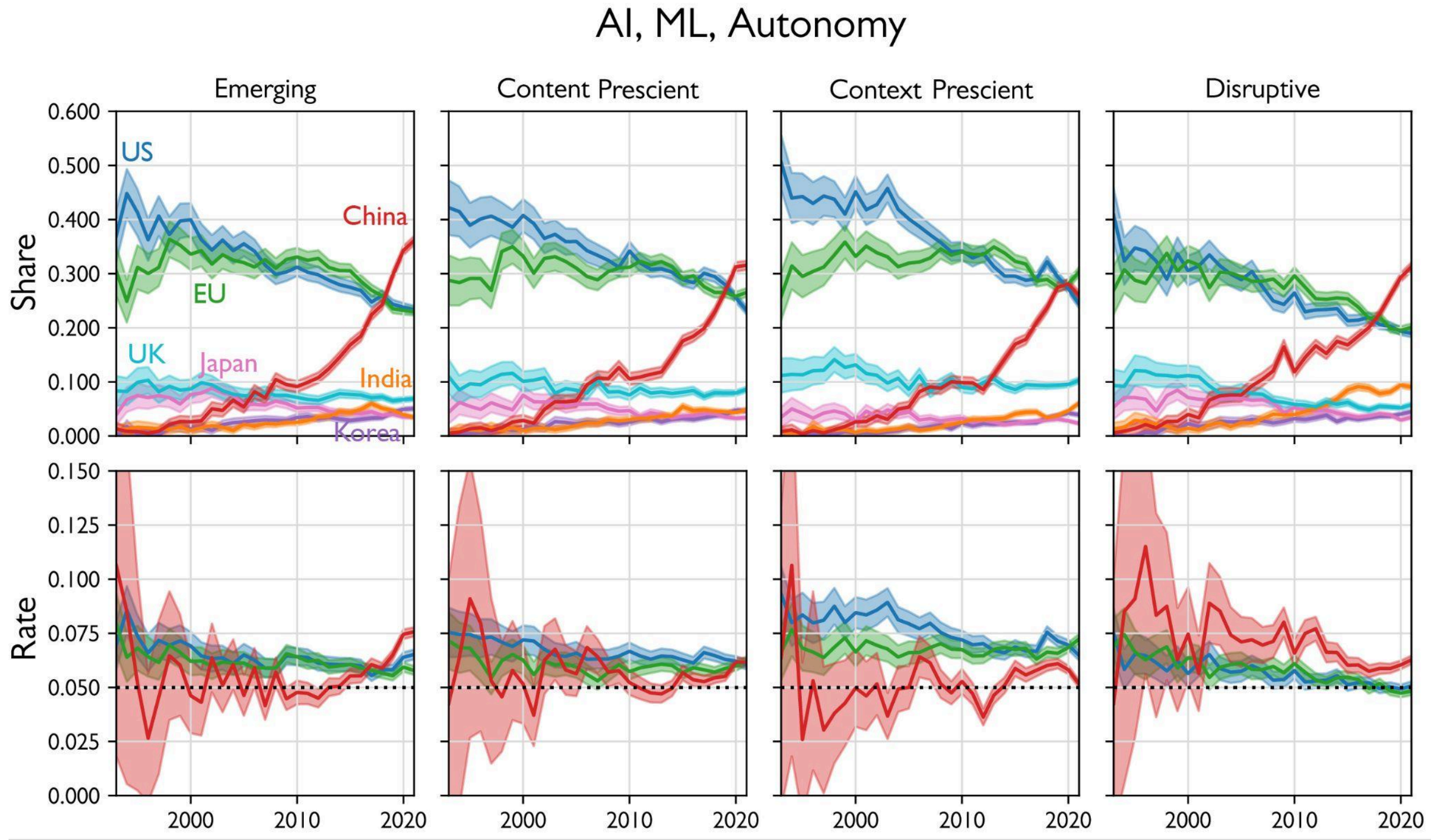

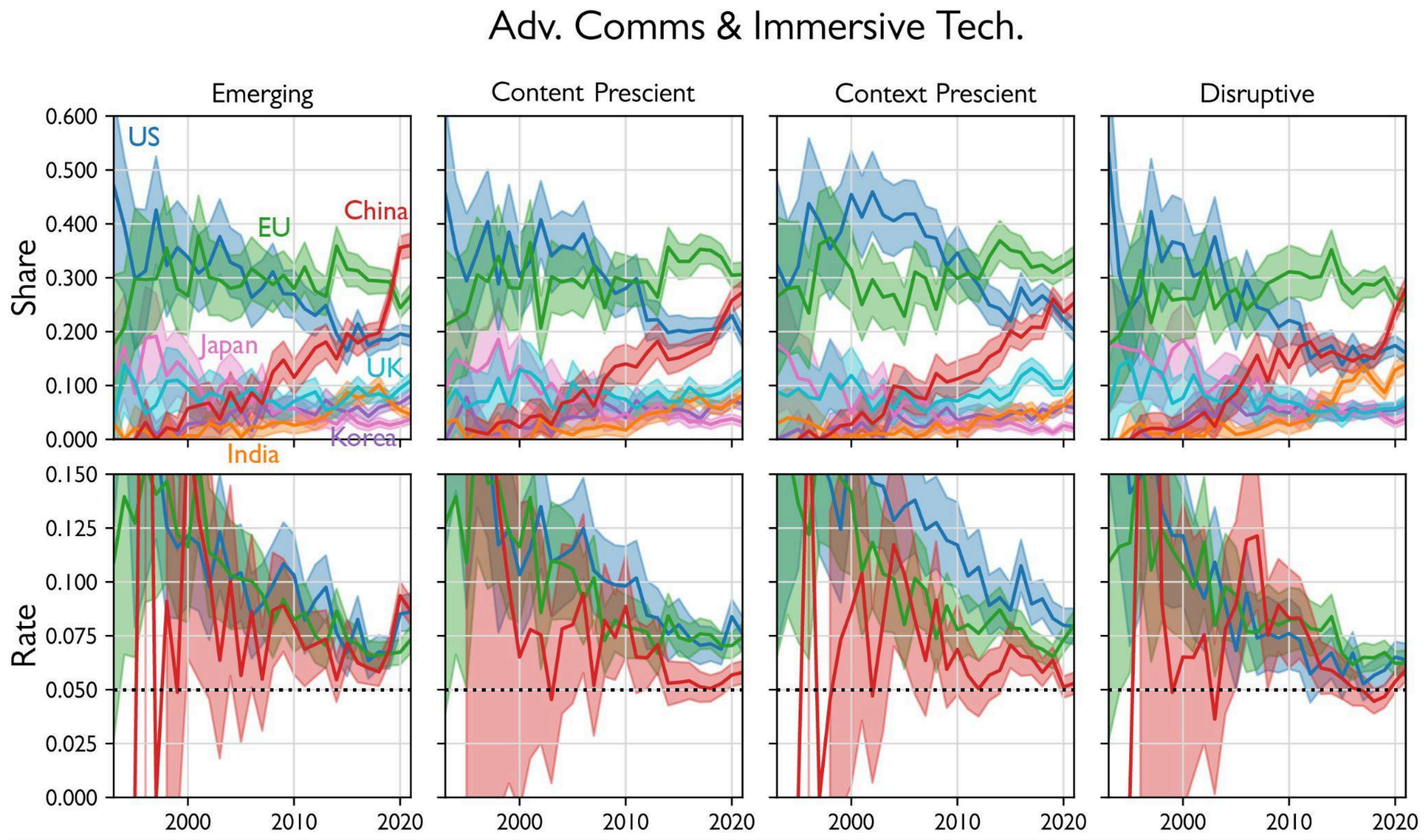

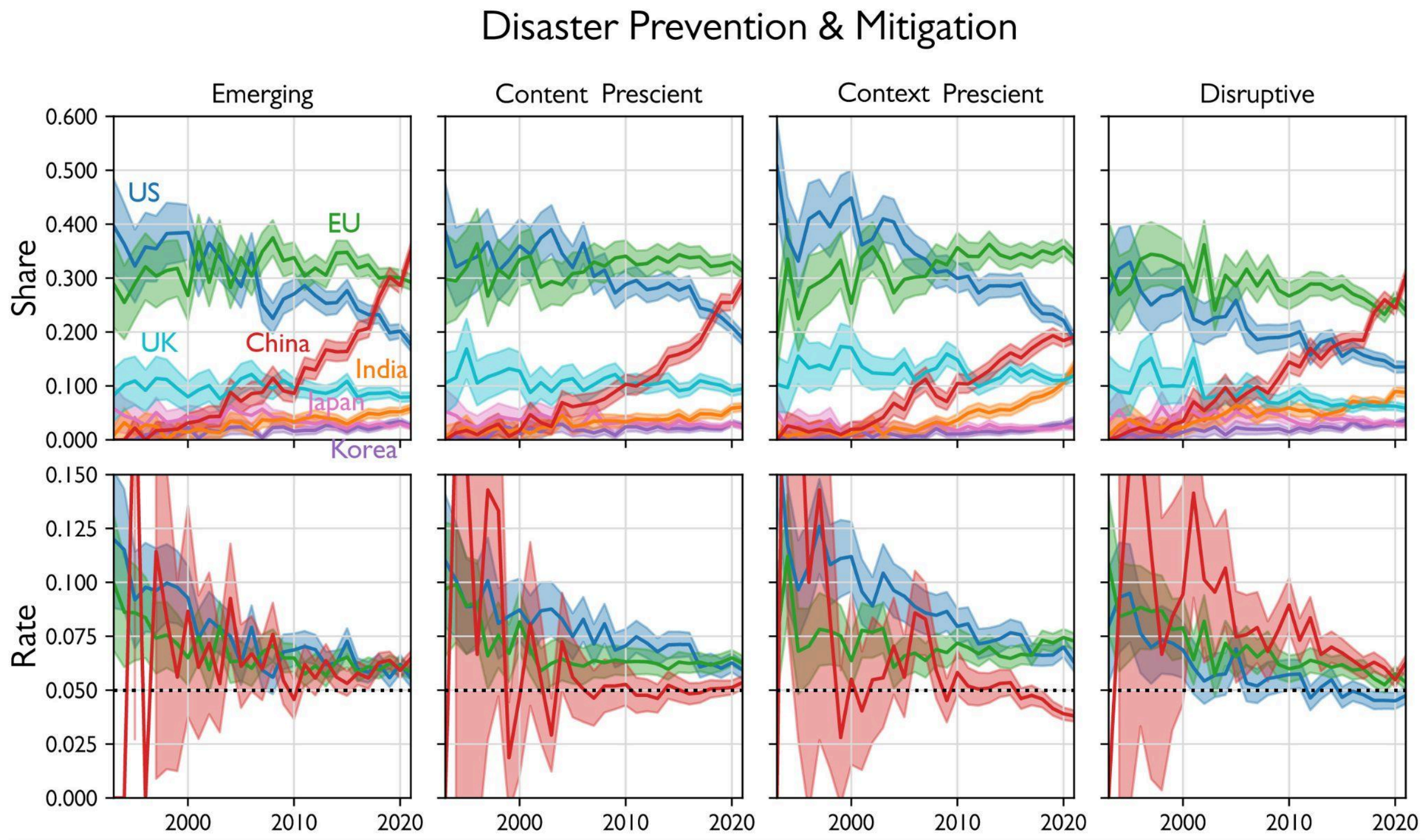

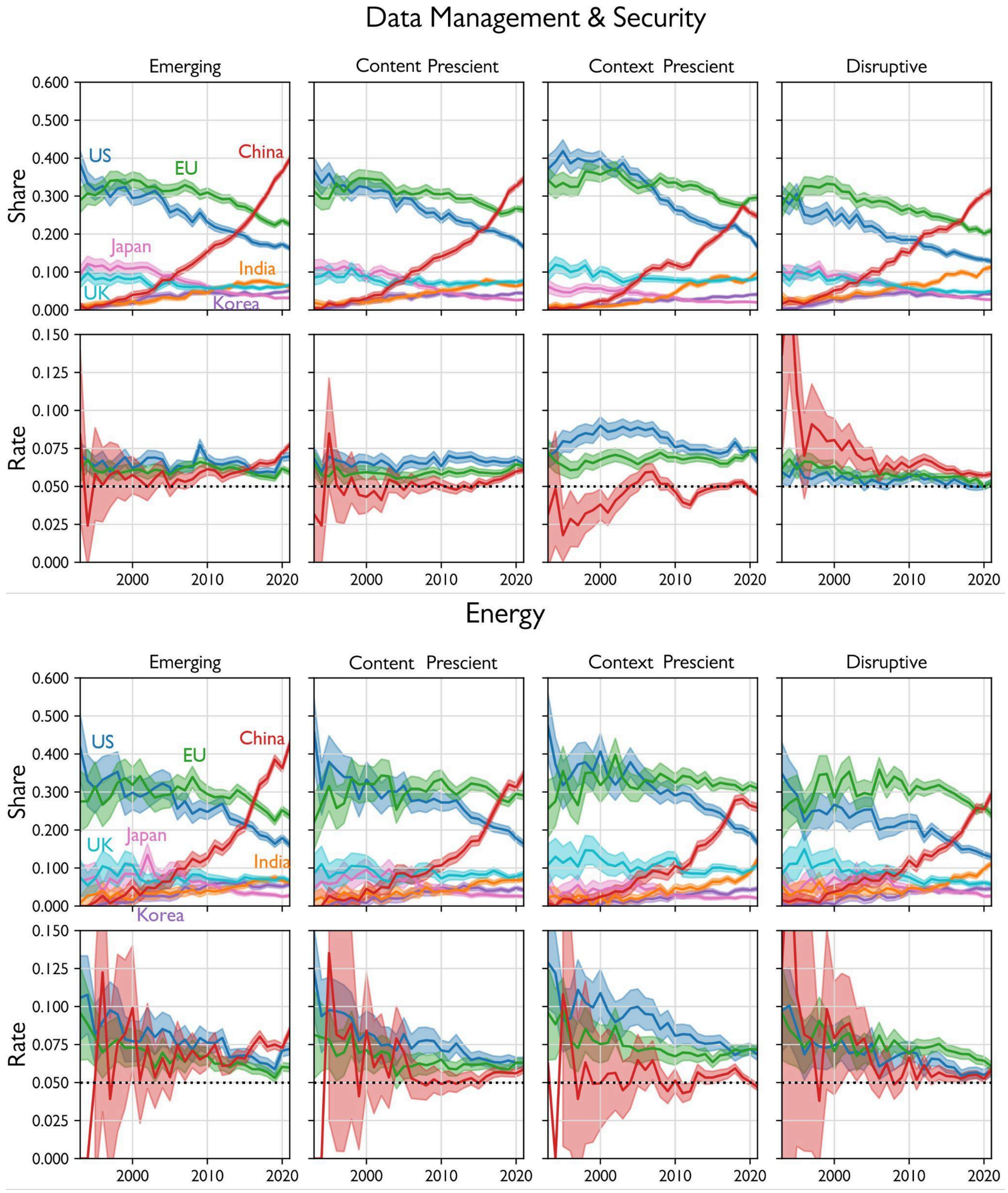

**Figure S8. Innovation patterns in National Science Foundation priority research areas.** Each panel shows the share of innovative papers (top plots) and rate of innovation relative to publication volume (bottom plots) for scientific fields associated with critical technology domains identified as strategically important by the NSF and U.S. Congress. Domains shown include: Biomedical Technology, Genomics, and Synthetic Biology; Artificial Intelligence,

Machine Learning, and Autonomy; Advanced Communications and Immersive Technology; Disaster Prevention and Mitigation; Data Management and Security; and Energy. Shaded regions indicate 95% confidence intervals. The 5% baseline rate is shown as a dashed horizontal line in rate plots. Results reveal domain-specific variation: China leads in emerging areas across most domains, while the US maintains advantages in prescience measures, particularly in AI/ML and biomedical fields. Confidence intervals are wider in smaller fields, reflecting greater uncertainty with fewer papers.

# References

1. D. O'Neil, English as the lingua franca of international publishing. *World Englishes* **37**, 146–165 (2018).

2. M. S. Di Bitetti, J. A. Ferreras, Publish (in English) or perish: The effect on citation rate of using languages other than English in scientific publications. *Ambio* **46**, 121–127 (2017).

3. B. Perozzi, R. Al-Rfou, S. Skiena, "DeepWalk: online learning of social representations" in *Proceedings of the 20th ACM SIGKDD International Conference on Knowledge Discovery and Data Mining* (Association for Computing Machinery, New York, NY, USA, 2014)*KDD '14*, pp. 701–710.

4. J. Sourati, J. Evans, Accelerating science with human-aware artificial intelligence, *arXiv [cs.AI]* (2023). http://arxiv.org/abs/2306.01495.